\documentclass[pra,aps,twocolumn,twoside,superscriptaddress]{revtex4}

\usepackage{amsmath,amsfonts,amssymb,color,graphics,graphicx,latexsym,revsymb,amsthm,url,verbatim,appendix,epstopdf}
\usepackage{algorithm}

\usepackage{algcompatible}

\usepackage{hyperref}

\usepackage{tikz}
\usetikzlibrary{decorations.shapes}
\usetikzlibrary{shapes.symbols}
\usetikzlibrary{decorations.pathmorphing}

\input{qcircuit}

\floatname{algorithm}{Protocol}

\newtheorem{theorem}{Theorem}
\newtheorem{corollary}[theorem]{Corollary}
\newtheorem{lemma}[theorem]{Lemma}
\newtheorem{proposition}[theorem]{Proposition}

\theoremstyle{definition}
\newtheorem{definition}{Definition}

\def\squareforqed{\hbox{\rlap{$\sqcap$}$\sqcup$}}
\def\qed{\ifmmode\squareforqed\else{\unskip\nobreak\hfil
\penalty50\hskip1em\null\nobreak\hfil\squareforqed
\parfillskip=0pt\finalhyphendemerits=0\endgraf}\fi}
\def\endenv{\ifmmode\;\else{\unskip\nobreak\hfil
\penalty50\hskip1em\null\nobreak\hfil\;
\parfillskip=0pt\finalhyphendemerits=0\endgraf}\fi}
\def\Dbar{\leavevmode\lower.6ex\hbox to 0pt
{\hskip-.23ex\accent"16\hss}D}

\makeatletter
\def\bcj{\begin{conjecture}}
\def\ecj{\end{conjecture}}
\def\bcr{\begin{corollary}}
\def\ecr{\end{corollary}}
\def\bd{\begin{definition}}
\def\ed{\end{definition}}
\def\bea{\begin{eqnarray}}
\def\eea{\end{eqnarray}}
\def\bem{\begin{enumerate}}
\def\eem{\end{enumerate}}
\def\bex{\begin{example}}
\def\eex{\end{example}}
\def\bim{\begin{itemize}}
\def\eim{\end{itemize}}
\def\bl{\begin{lemma}}
\def\el{\end{lemma}}
\def\bpf{\begin{proof}}
\def\epf{\end{proof}}
\def\bpp{\begin{proposition}}
\def\epp{\end{proposition}}
\def\bqu{\begin{question}}
\def\equ{\end{question}}
\def\br{\begin{remark}}
\def\er{\end{remark}}
\def\bt{\begin{theorem}}
\def\et{\end{theorem}}

\def\btb{\begin{tabular}}
\def\etb{\end{tabular}}

\newcommand{\nc}{\newcommand}

 \nc{\bA}{{\bf A}} \nc{\bB}{{\bf B}} \nc{\bC}{{\bf C}}
 \nc{\bD}{{\bf D}} \nc{\bE}{{\bf E}} \nc{\bF}{{\bf F}}
 \nc{\bG}{{\bf G}} \nc{\bH}{{\bf H}} \nc{\bI}{{\bf I}}
 \nc{\bJ}{{\bf J}} \nc{\bK}{{\bf K}} \nc{\bL}{{\bf L}}
 \nc{\bM}{{\bf M}} \nc{\bN}{{\bf N}} \nc{\bO}{{\bf O}}
 \nc{\bP}{{\bf P}} \nc{\bQ}{{\bf Q}} \nc{\bR}{{\bf R}}
 \nc{\bS}{{\bf S}} \nc{\bT}{{\bf T}} \nc{\bU}{{\bf U}}
 \nc{\bV}{{\bf V}} \nc{\bW}{{\bf W}} \nc{\bX}{{\bf X}}
 \nc{\bZ}{{\bf Z}}

\nc{\cA}{{\cal A}} \nc{\cB}{{\cal B}} \nc{\cC}{{\cal C}}
\nc{\cD}{{\cal D}} \nc{\cE}{{\cal E}} \nc{\cF}{{\cal F}}
\nc{\cG}{{\cal G}} \nc{\cH}{{\cal H}} \nc{\cI}{{\cal I}}
\nc{\cJ}{{\cal J}} \nc{\cK}{{\cal K}} \nc{\cL}{{\cal L}}
\nc{\cM}{{\cal M}} \nc{\cN}{{\cal N}} \nc{\cO}{{\cal O}}
\nc{\cP}{{\cal P}} \nc{\cQ}{{\cal Q}} \nc{\cR}{{\cal R}}
\nc{\cS}{{\cal S}} \nc{\cT}{{\cal T}} \nc{\cU}{{\cal U}}
\nc{\cV}{{\cal V}} \nc{\cW}{{\cal W}} \nc{\cX}{{\cal X}}
\nc{\cZ}{{\cal Z}}

\nc{\hA}{{\hat{A}}} \nc{\hB}{{\hat{B}}} \nc{\hC}{{\hat{C}}}
\nc{\hD}{{\hat{D}}} \nc{\hE}{{\hat{E}}} \nc{\hF}{{\hat{F}}}
\nc{\hG}{{\hat{G}}} \nc{\hH}{{\hat{H}}} \nc{\hI}{{\hat{I}}}
\nc{\hJ}{{\hat{J}}} \nc{\hK}{{\hat{K}}} \nc{\hL}{{\hat{L}}}
\nc{\hM}{{\hat{M}}} \nc{\hN}{{\hat{N}}} \nc{\hO}{{\hat{O}}}
\nc{\hP}{{\hat{P}}} \nc{\hR}{{\hat{R}}} \nc{\hS}{{\hat{S}}}
\nc{\hT}{{\hat{T}}} \nc{\hU}{{\hat{U}}} \nc{\hV}{{\hat{V}}}
\nc{\hW}{{\hat{W}}} \nc{\hX}{{\hat{X}}} \nc{\hZ}{{\hat{Z}}}

\nc{\hn}{{\hat{n}}}

\def\diag{\mathop{\rm diag}}

\def\ox{\otimes}

\newcommand{\ket}[1]{|#1\rangle}

\newcommand{\ketbra}[2]{|#1\rangle\!\langle#2|}

\newcommand{\tgate}{{\sf T}}

\newcommand{\pgate}{{\sf P}}

\newcommand{\cnot}{{\sf CNOT}}

\begin{document}
\title{Instantaneous nonlocal quantum computation and circuit depth reduction}
\author{Li Yu}\email{yuli@hznu.edu.cn}
\affiliation{School of Physics, Hangzhou Normal University, Hangzhou, Zhejiang 311121, China}
\author{Jie Xu}
\affiliation{School of Physics, Hangzhou Normal University, Hangzhou, Zhejiang 311121, China}
\author{Fuqun Wang}
\affiliation{School of Mathematics, Hangzhou Normal University, Hangzhou, Zhejiang 311121, China}
\affiliation{Key Laboratory of Cryptography of Zhejiang Province, Hangzhou 311121, China}
\affiliation{Westone Cryptologic Research Center, Beijing 100071, China}
\author{Chui-Ping Yang}\email{yangcp@hznu.edu.cn}
\affiliation{School of Physics, Hangzhou Normal University, Hangzhou, Zhejiang 311121, China}

\begin{abstract}
Instantaneous two-party quantum computation is a computation process with bipartite input and output, in which there are initial shared entanglement, and the nonlocal interactions are limited to simultaneous classical communication in both directions. It is almost equivalent to the problem of instantaneous measurements, and is related to some topics in quantum foundations and position-based quantum cryptography. In the first part of this work, we show that a particular simplified subprocedure, known as a garden-hose gadget, cannot significantly reduce the entanglement cost in instantaneous two-party quantum computation. In the second part, we show that any unitary circuit consisting of layers of Clifford gates and T gates can be implemented using a circuit with measurements (or a unitary circuit) of depth proportional to the T-depth of the original circuit. This result has some similarity with and also some difference from a result in measurement-based quantum computation. It is of limited use since interesting quantum algorithms often require a high ratio of T gates, but still we discuss its extensions and applications.
\end{abstract}
\maketitle


\section{Introduction}\label{sec1}

Instantaneous two-party quantum computation is a computation process with bipartite input and output, in which Alice and Bob initially share some entanglement, and the nonlocal interactions are limited to simultaneous classical communication in both directions. (More details are given in Sec.~\ref{sec2}.) This problem is almost equivalent to the problem of instantaneous measurements, and is thus related to some topics in quantum foundations. These two problems are central for position verification protocols in position-based quantum cryptography \cite{Malaney10,BCF11,LL11,KMS11,Beigi11,CL15,BC16,Malaney16a,RTK18,XPW19,Li19,BCS22}. There are some instantaneous quantum computation protocols in the literature (sometimes called ``fast unitary protocols''), but they are either for specific classes of circuits \cite{ygc12,Yu11}, or have large entanglement cost \cite{CCJ10,BCF11,Beigi11,Speelman16,GC20}, usually exponential in the $T$-depth of the circuit.

Many instantaneous quantum computation protocols in the literature use a type of entangled resource state with associated operations, called garden-hose gadget \cite{BFSS13}. In the first part of this work, we show that the use of a particular type of garden-hose gadget that appeared in the study of quantum homomorphic encryption (QHE) cannot significantly reduce the entanglement cost in generic instantaneous quantum computation. The form of the gadget is in \cite{Yu18}, which is a simplified version of an example gadget in \cite{Dulek16}, but the purpose of the examples in \cite{Yu18} or \cite{Dulek16} is not for instantaneous computation or measurement.

Independent from the first part of this work, we also show that any unitary circuit consisting of Clifford stages and T gates can be implemented using a circuit with measurements (or a unitary circuit) of depth only proportional to the T-depth of the original circuit. The size of the modified circuit is roughly of the same order as the original circuit. There is a similar result in measurement-based quantum computation, which says a measurement-based computation with only Clifford gates can be implemented in one step \cite{RBB03}. But the result here can get rid of the need of measurement on many qubits by using some modest time cost. This result is of limited use since existing quantum algorithms often require the $T$ gates to be densely distributed in the circuit, but still we discuss some extensions and applications of the result.

The two parts of this work concern the nonlocal depth of a two-party protocol, and the local depth of a circuit, respectively. And the underlying techniques are somewhat similar since they both use teleportations. We hope more connections of the two problems can be found.

The rest of the paper is organized as follows. In Sec.~\ref{sec2} we introduce some background knowledge. In Sec.~\ref{sec3} we discuss the use of a simplified garden-hose gadget in instantaneous quantum computation. In Sec.~\ref{sec4}, we show the reduction of the depth of a unitary circuit by running the Clifford parts in parallel, and describe some extensions and applications. The Sec.~\ref{sec5} contains the conclusion and some open questions.

\section{Preliminaries}\label{sec2}

Firstly, we describe what two-party instantaneous quantum computation is. In such computation, a unitary is performed on data provided by both parties. The unitary is known to both parties, and each party's input is some quantum state, not necessarily known to the other party. The two party's input states are usually not entangled with each other. The two parties also share some initially entangled state, which is the resource to be used in the protocol. The computation process is such that the interactions between the two parties are limited to simultaneous classical communication in both directions. The two parties can do any local operations, before and after the message exchange. The local operations are regarded as fast, i.e. their time cost is negligible compared to the time cost of communication. The time for preparing the initial entanglement is ignored. This means the whole process is regarded as requiring time cost equal to the time cost of one-way communication. This does not sound ``instantaneous''; the word ``instantaneous'' has its origin in the study of instantaneous measurements. In instantaneous measurements, the two parties perform some local measurements (with the help of shared entangled states), and they exchange the local measurement outcomes and perform some classical post-processing. In some sense the measurement is fixed (modulo post-processing) when they perform the local measurements, hence the name ``instantaneous''.

Some notations are as follows. Denote $\ket{+}:=\frac{1}{\sqrt{2}}(\ket{0}+\ket{1})$, and $\ket{-}:=\frac{1}{\sqrt{2}}(\ket{0}-\ket{1})$. Let $I=\diag(1,1)$, $Z=\diag(1,-1)$, $X=\begin{pmatrix}1 & 0 \\ 0 & 1 \end{pmatrix}$, $H=\frac{1}{\sqrt{2}}\begin{pmatrix}1 & 1 \\ 1 & -1 \end{pmatrix}$, $P=\diag(1,i)$, $\cnot=\ketbra{0}{0}\ox I+\ketbra{1}{1}\ox X$, $T=\diag(1,e^{i\pi/4})$, and $R_z(\theta)=\diag(1,e^{i\theta})$. When the symbol $\oplus$ is used between numbers, it represents addition modulo 2.

The qubit Pauli group is the group generated by $X$ and $Z$. We call $I$, $X$, $Z$ and $XZ$ as the Pauli operators. The $n$-qubit Clifford group is generated by $H$, $P$ and $CNOT$ (each may act on different qubits), and the members of such group, with overall phase ignored, are called Clifford operators. When the Pauli operators commute through the Clifford operators, they become Pauli operators: 
\begin{eqnarray}\label{eq:keyupdate1}
&&P X=i X Z P,\quad\quad P Z=Z P,\notag\\
&&H X=Z H,\quad\quad\quad H Z=X H,\notag\\
&&\cnot_{12} (X_1^a Z_1^b \ox X_2^c Z_2^d)= (X_1^a Z_1^{b\oplus d} \ox X_2^{a\oplus c} Z_2^d)\cnot_{12},\notag\\
\end{eqnarray}
and from these equations we find that for an initial Pauli operator of the form $X^a Z^b$, the exponents $a$ and $b$ would undergo a linear map under commutation through the Clifford operators. This linearity property means that contributions to $a$ and $b$ from different sources (Alice and Bob) can be transformed independently under commutation through the Clifford operators.

The $T=\diag(1,e^{i\pi/4})$ is not in the Clifford group. When the Pauli operators $X$ and $Z$ are commuted through the $T$ gate, the following relation holds:
\begin{eqnarray}\label{eq:keyupdate2}
Z T &=& T Z,\\
T X &=& e^{-i\pi/4} P X T.
\end{eqnarray}
The similar commutation rule for the Pauli operator $XZ$ can be derived from the above. In this paper we always write Pauli operators in the form of $X^a Z^b$ up to phases, thus we do not have to be concerned about the commutation rule for $XZ$. If we ignore the $P$ and the overall phase, we see that the exponents for $X$ and $Z$ are both unchanged under the $T$.

Next, we introduce the main idea of the computationally secure QHE scheme by Dulek et al \cite{Dulek16}. The main idea is that Alice hides the initial Pauli masks from her initial teleportation, and only sends Bob the encrypted values. The necessary $P^\dag$ corrections after Bob's $T$ gates due to such mask and both parties' subsequent measurement outcomes are compensated by Alice's and Bob's joint operations in the garden-hose gadgets. 

\section{On the possible use of a gadget in instantaneous two-party quantum computation}\label{sec3}

A generic quantum protocol for instantaneous two-party quantum computation is presented as Protocol~\ref{ptl1} below. The main idea is that Alice delays sending the initial Pauli masks from her initial teleportation, and she keeps Pauli masks to herself during the execution of the circuit, with the necessary $P^\dag$ correction due to such mask and her subsequent measurement outcomes compensated by her operations in the garden-hose gadgets. Each garden-hose gadget connects a Clifford+T stage of Bob's circuit to the next stage, while implementing a possible $P^\dag$ correction. The correction is expressed as $(P^\dag)^g$, where the exponent $g$ is generally a function of the two parties' measurement results including Alice's initial Pauli masks.

\begin{algorithm*}[htb]
\caption{A generic instantaneous two-party quantum computation protocol based on garden-hose gadgets.}\label{ptl1}
\begin{flushleft}
\noindent{\bf Input:} Alice and Bob have quantum registers in the joint state $\ket{\Psi}$, and they share a maximally entangled resource state $\ket{\Phi}$. The unitary operator $U$ to be performed is known to both parties. A decomposition of the $U$ in terms of Clifford gates and $T$ gates is also known to both parties.\\
\noindent{\bf Output:} The output is a two-party quantum state $U\ket{\Psi}$.\\
\begin{enumerate}
\item Each party selects the local qubits in the qubits for $\ket{\Phi}$ using an agreed-upon plan, so that each party knows which local qubits are for which garden-hose gadget associated with a $T$ gate.
\item Alice performs the local measurements in an initial teleportation step, but does not tell the measurement results to Bob until a predetermined time after she completes her other operations in the garden-hose gadgets. She records the initial outcomes and the measurement outcomes in subsequent garden-hose gadgets, to update her later choice of inputs in the garden-hose gadgets. She proceeds through all the garden-hose gadgets, effectively reaching the end of the circuit.
\item (This step is done simultaneously with the previous step.) Bob performs the local Clifford gates, $T$ gates and measurements in the garden-hose gadgets and teleportations, and after each layer of $T$ gates and the subsequent measurements, he records the outcomes locally, in order to update his input in the subsequent garden-hose gadgets. He proceeds through the end of the circuit. Bob performs the measurements in the teleportation of the  part of the output qubits which are to be sent to Alice. 
\item The two parties simultaneously initiate the classical communication of the outcomes of the previous Bell-state measurements.
\item Alice and Bob each does local Pauli corrections according to the received message, to obtain the local part of the final output.\\
\end{enumerate}
\end{flushleft}
\end{algorithm*}

In general a protocol may need different garden-hose gadgets at different $T$ gates. The particular garden-hose gadget we consider here is the same as in \cite{Yu18}, which is a simplified version of a gadget in \cite{Dulek16} . For completeness, we include the diagram for the gadget here. The Fig.~\ref{fig:toy2} below shows a simplified version of a gadget in \cite{Dulek16} for correcting an unwanted $\pgate$ gate due to a $\tgate$ gate in the circuit with certain prior Pauli corrections. The input qubit starts from the position ``in'', and ends up in a qubit on Bob's side labeled ``out1'' or ``out2'', depending on Bob's input bit $p$. The unwanted $\pgate$ on this qubit is corrected, but some other Pauli corrections now arise from the Bell-state measurements. These Pauli corrections are recorded by the local party, and would affect each party's local input in the later garden-hose gadgets. But the later garden-hose gadgets have to be different from the current one, due to that the required corrections is related to the measurement outcomes in a complex way: some Pauli corrections which depend on the product of bits from Alice and Bob are now need. We can find two types of these terms: one type is for the Pauli masks due to the initial Bell-state measurement on Bob's side in the gadget, which would need a correction when they pass through the possible $P^\dag$ gate on Alice's side. The other is for Alice's Pauli masks for the measurement by Alice on the unused two qubits; the choice of which qubits are ``unused'' is determined by Bob.

Due to these terms that depend on both parties' input, we will need larger garden-hose gadgets for the subsequent $T$ gates in the circuit to be evaluated correctly. According to \cite{Speelman16}, the required size of the garden-hose gadgets in later part of the circuit is expected to increase exponentially as the $T$-depth increases. Hence, the use of our particular gadget in some initial stage does not qualitatively change the exponential entanglement cost of instantaneous nonlocal quantum computation.

Here is a note on the issue of the output being possibly on different qubits: this is acceptable if Bob has not done later computations. But if Bob has effectively performed some steps of the later computations, such as in the method in the next section, he could use an additional teleportation with selective input qubit to bridge the output of the current garden-hose gadget with his subsequent computations.

\begin{figure*}[htb]
\centering
\begin{tikzpicture}[decoration=snake]
\filldraw (0,0) circle (2pt);
\filldraw (2,0) circle (2pt);
\draw[decorate] (0,0) -- (2,0);

\filldraw (0,-1) circle (2pt);
\filldraw (2,-1) circle (2pt);
\draw[decorate] (0,-1) -- (2,-1);

\filldraw (0,1) circle (2pt);
\filldraw (2,1) circle (2pt);
\draw[decorate] (0,1) -- (2,1);

\filldraw (0,2) circle (2pt);
\filldraw (2,2) circle (2pt);
\draw[decorate] (0,2) -- (2,2);

\node at (-1.5,4) {$p$};
\node at (-3,3.5) {$0$};
\node at (-0.3,3.5) {$1$};
\draw (-0.3,1) to [bend left=50] (-0.3,3);
\draw (-3,2) to [bend left=50] (-3,3);

\node at (5,4) {$q$};
\node at (3.3,3.5) {$0$};
\node at (6.5,3.5) {$1$};
\draw (3.3,0) to [bend right=50] (3.3,2);
\draw (6.5,-1) to [bend right=50] (6.5,1);

\filldraw (0,3) circle (2pt);
\node[anchor=west] at (0,3) {in};

\node[anchor=west] at (-0.2,0.3) {out1};
\node[anchor=west] at (-0.2,-0.7) {out2};

\draw (3.3,-1) to [bend right=50] (3.3,1) to (2.3,1);
\draw (6.5,0) to [bend right=50] (6.5,2) to (5.5,2);
\filldraw[fill=white] (5.7,1.75) rectangle (6.3,2.25);
\filldraw[fill=white] (2.5,0.75) rectangle (3.1, 1.25);
\node at (6,2) {$\pgate^\dag$};
\node at (2.8,1) {$\pgate^\dag$};

\draw[dashed] (-4.5,-2) rectangle (0.75,5);
\draw[dashed] (1.3,-2) rectangle (8,5);
\node[anchor=south] at (-4,5) {Bob};
\node[anchor=south] at (7.5,5) {Alice};
\end{tikzpicture}
\caption{A simplified version of a gadget in \cite{Dulek16} for applying a $\pgate^\dag$ to a qubit initially at the position ``in'' if and only if $p+q=1\,\,(\rm{mod}\,\,2)$, using the ``garden hose'' method. The dots connected by wavy lines are EPR pairs. The curved lines are for Bell-state measurements. For example, if $p=0$ and $q=1$, the qubit is teleported through the first and the third EPR pairs, with a $\pgate^\dag$ applied to it by Alice in between. The state of the input qubit always ends up in a qubit on Bob's side, and the position depends on Bob's input bit $p$: if $p=0$, the output is on the qubit labeled ``out1'', otherwise it is on the qubit labeled ``out2''.}
\label{fig:toy2}
\end{figure*}

The local computations are still sequential on both parties. The savings in time is only in that the two directions of classical communication is done simultaneously. This fits the definition of instantaneous quantum computation. When the local gates are regarded as fast (costing no time), then the overall time cost is equal to the time for one-way communication from Alice to Bob. 

Since the final corrections are Pauli operators, for obtaining classical outcomes, the final local measurement bases can be viewed as fixed, i.e. independent of the final corrections sent by the other party. Hence, adding local measurements to the Protocol~\ref{ptl1} would yield a protocol for instantaneous two-party measurement.

\section{Circuit depth reduction of unitary circuits using teleportation}\label{sec4}

\begin{algorithm*}[htb]
\caption{Conversion of a generic unitary circuit composed of Clifford and $T$ stages to a circuit with intermediate measurements.}\label{ptl2}
\begin{flushleft}
\noindent{\bf Input:} A unitary circuit with explicit decomposition, consisting of interleaving layers of Clifford circuits and $T$ gates.\\
\noindent{\bf Output:} A circuit consisting of unitary gates and intermediate measurements, with some initial entangled state as a resource.\\
\begin{enumerate}
\item Suppose the circuit is on $n$ qubits, and has $K$ layers of Clifford gates followed by $T$ gates. Firstly, prepare $n(K-1)$ pairs of qubits. Execute each of the later $K-1$ Clifford+T sub-circuits in the original circuit on the second qubits on each of the $n$ pairs for such sub-circuit, except for the first Clifford+T sub-circuit. The first Clifford+T sub-circuit is evaluated on the original input qubits.
\item Let the initial Pauli masks ($2n$ bits) be all zero. For each layer of Clifford and $T$ gates in the original circuit, perform appropriate $P^\dag$ correction(s) on the output qubits of such sub-circuit according to the current Pauli mask, and perform measurements in a teleportation (involving measuring the current output qubit and the first qubit in some previously prepared qubit pairs) towards a qubit for the next Clifford sub-circuit, and record the measurement outcomes, and update the current Pauli mask.
\item At the end of the circuit, correct the result using the current Pauli mask.\\
\end{enumerate}
\end{flushleft}
\end{algorithm*}

The protocol is shown as Protocol~\ref{ptl2}. Its idea is simple: firstly prepare some entangled pairs. Then, complete the Clifford+T sub-circuits on distinct qubits, where the input qubits for each sub-circuit are the second qubits in the prepared pairs, so they are entangled with the first qubits in the pairs (except for the first Clifford+T sub-circuit). And then perform teleportations to link the different sub-circuits, while performing appropriate $P^\dag$ corrections before the measurements in teleportations. The linking process has to be sequential, thus the total depth of the new circuit (with intermediate measurements) is proportional to the $T$-depth of the original circuit, plus the depth of a Clifford computation (the latter may be reduced to a constant if we apply the operations in measurement-based computation for implementing the Clifford circuits \cite{RBB03}). It is not hard to turn this circuit into a fully unitary circuit.

A simple calculation for verifying this protocol can be done with one qubit as the example case: suppose qubits $s$ and $t$ are initially in the entangled state $\frac{1}{\sqrt{2}}(\ket{00}+\ket{11})$. Let qubit $r$ to be in the state $\alpha\ket{0}+\beta\ket{1}$, which may be the result of an initial stage of the overall circuit. Perform a unitary operator $U$ on qubit $t$. Next, we perform the teleportation, i.e. perform Bell-state measurement on qubits $r$ and $s$, and according to the outcome (2 bits), correct the state of $t$ using Pauli operators. The final state of $t$ would become $U X^a Z^b(\alpha\ket{0}+\beta\ket{1})$. This differs from the desired form in that the Pauli operators are applied prior to $U$. To commute the Pauli operators to after $U$, we take advantage of the fact that each sub-circuit in the Protocol~\ref{ptl2} is a Clifford circuit followed by a layer of $T$ gates, and regard such sub-circuit as the $U$ here. Thus we can always map the initial Pauli masks to some final Pauli masks, with possible $P^\dag$ corrections after the $T$ gates. 

The example above also helps to explain why we cannot pre-compute two Clifford+T stages at a time in Protocol~\ref{ptl2}; but the exact reason can be traced back to that the circuits with two layers of $T$ gates (such as $T H T$) cannot be generally reduced to a layer of $T$ gates with the help of Clifford gates \cite{Selinger13}.

This result is somewhat similar to a result in measurement-based quantum computation, which says a measurement-based computation with only Clifford gates can be implemented in one step, and we can remove from each quantum algorithm (in measurement-based computation) its Clifford part \cite{RBB03}. But a difference is in the space cost: for the current protocol, if we do not implement the Clifford sub-circuits in the measurement-based way, we can save the qubits needed in measurement-based computation of such Clifford circuit; and the cost is just some more time to implement the Clifford circuits in parallel, which is not much considering that it often depends linearly on the number of qubits $n$ \cite{Maslov07}, while the overall circuit depth is often much larger than $n$.

Since $T$ gates are often dense in some useful quantum algorithms, the plain use of the Protocol~\ref{ptl2} is of limited utility. An extension of the protocol for the case of classical reversible circuits is as follows: evaluate some Clifford+T stages at once, with guessed $P^\dag$ corrections in between the stages. In this way, many partial circuits is pre-generated, and some of them may be linked later, while some of them may be discarded later. For this to be possible, the inputs to these sub-circuits need to be supplied, so that we require such input to be classical, for it to be copied many times. This is the reason why we require the circuit to be a classical reversible circuit (with classical input) here.

If we link $r$ stages at a time, we need about $O(2^r)$ copies of such sub-circuit, to let one of them have the actually needed $P^\dag$ corrections. Thus this approach accelerates the overall time by a factor about $r$, while expanding the space requirement by a factor of about $O(2^r)$.

For quantum circuits that are with non-classical state as input state for each sub-circuit, the above method does not work, due to that the Pauli masks from the last teleportation to link qubits can affect the correctness of the guessed $P^\dag$ corrections in the precomputed circuit. But the plain use of the result (evaluating one Clifford+T layer at a time) is already quite good for the case that there are not too many $T$ gates in a circuit. A possible application is to study the effect of randomly added phase noise (related to $T$ gates but not necessarily equivalent) in a Clifford dynamics, by quantum simulation.

We can think of another extension of the result to the Clifford + cyclotomic gate set \cite{FDV15}, instead of the Clifford + T gate set. The extension of the protocol to this case seems easy, so we omit it here.

\section{Conclusion}\label{sec5}

In this work we have discussed the possible use of a simplified garden-hose gadget in instantaneous two-party quantum computation, showing that it cannot significantly reduce the entanglement cost. We also presented a protocol for reducing the time cost of evaluating the Clifford part of the circuit for a generic unitary circuit.

An interesting problem is the whether the protocols or techniques here can help secure two-party computation in general. We think that secure evaluation of quantum programs which have non-orthogonal quantum outputs for different classical inputs may be an interesting problem, since for the case of orthogonal output, the known no-go results (e.g. \cite{Lo97,bcs12}) for two-party evaluation of classical functions seem to put constraints on the type of security that might be achieved.

\smallskip
\section*{Acknowledgments}

This research is supported by the National Natural Science Foundation of China (No. 11974096, No. 61972124, No. 11774076, and No. U21A20436), and the NKRDP of China (No. 2016YFA0301802).

\linespread{1.0}
\bibliographystyle{unsrt}
\bibliography{homo}

\end{document}